\documentclass[12pt]{article}
\usepackage{times}
\usepackage{geometry}
\usepackage{graphicx}
\usepackage[dvipsnames]{xcolor}
\usepackage[utf8]{inputenc}
\usepackage{enumitem,amssymb}
\usepackage{natbib}

\usepackage{ragged2e}
\usepackage{hyperref}
\hypersetup{
    colorlinks=true,
    citecolor=black,
    linkcolor=blue,
    filecolor=magenta,      
    urlcolor=ForestGreen,
}
\newlist{thematic}{itemize}{8}
\setlist[thematic]{label=$\square$}
\usepackage{pifont}

\begin{document}
\raggedright
\huge
Pathways to Sustainable Planetary Science \linebreak
\large
Planetary Science and Astrobiology Decadal Survey 2023-2032 White Paper \linebreak
\normalsize

\begin{figure}[ht]
    \centering
    \includegraphics[width=0.67\textwidth]{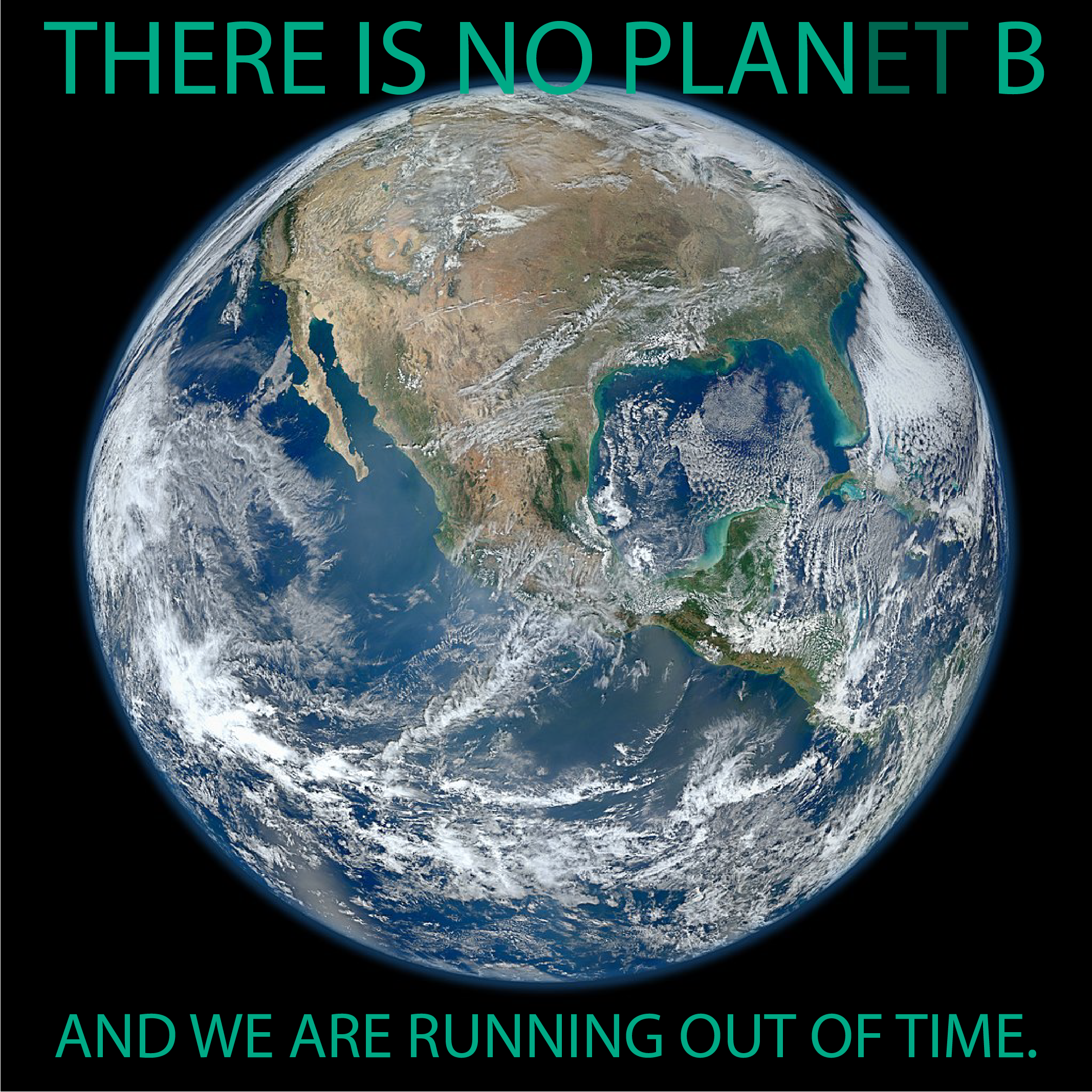}
\end{figure}
  
\textbf{Principal Author:}

Name: Matija \'Cuk	
 \linebreak						
Institution:  DPS Environmental Affairs Subcommittee, SETI Institute
 \linebreak
Email: mcuk@seti.org
 \linebreak
Phone: 1-650-810-0210 
 \linebreak
 
\textbf{Co-authors:} 
Anne K. Virkki (Arecibo Observatory/University of Central Florida)
Tom\'a\v s Kohout (University of Helsinki)
Emmanuel Lellouch (Observatoire de Paris)
Jack J. Lissauer (NASA Ames)
  \linebreak
  
\textbf{Endorsers:} Endorse the paper here: \url{https://forms.gle/iCMd1SKaQajvGU8w5}. 

\pagebreak
\section{Executive Summary}

Climate change is a major impending threat to the future of humanity \citep{nas14}. According to the International Panel on Climate Change (IPCC), our emissions are estimated to have caused 0.8$^\circ$C-1.2$^\circ$C of anthropogenic global warming (AGW) above pre-industrial levels \citep{ipcc}. AGW is likely to reach 1.5$^\circ$C between 2030 and 2052 if it continues to increase at the current rate. As the climate change is driven by the release of carbon dioxide and other greenhouse gases (GHG) into the atmosphere, there is a broad consensus that the mitigation of climate change requires transition to low GHG emission energy sources, technologies and practices. Implementing such changes systematically from individual to community-wide scales together with the resulting cultural changes and leadership towards environmental consciousness and responsibility are crucial to mitigate the looming damage of AGW. 

\

Given planetary scientists' wide recognition of the realities of climate change, and the need for us to maintain credibility by leading by example, it is appropriate to make own professional behavior more environmentally responsible. While  scientists  are  few in numbers, and planetary scientists far fewer, high volumes of academic travel to conferences, panels, colloquia, and research collaboration visits together with extensive use of large, energetically demanding infrastructures make the ``carbon footprint'' of scientists much higher than that of an average citizen. 


\

This White Paper focuses on how modifying our activities, particularly  associated with academic travel, can affect the carbon footprint of the planetary science community, and it makes recommendations on how the community and the funding agencies could best participate in the cultural change required to mitigate the damage that AGW will cause \citep[e.g., ][]{mat19}. 

\vspace{-.1cm}
\section{Conferences}

\citet{klo20} recently conducted an analysis of the carbon footprint of the 2019 American Geophysical Union (AGU) Fall meeting, which provides an apt example of the environmental effects of scientific meetings. Unsurprisingly, most of the estimated GHG emission caused by the AGU are due to air travel, with the outsized contribution by attendees from outside North America. In-state attendees have contributed a relatively low fraction of GHG emissions, by taking trains or buses, driving, or even short-haul flights (the former have much lower per-mile emissions, but the end total is not effected much by the choices of ``local'' attendees). While the details depend on exact technological and organizational solutions, there are several broad ways to move forward with more environmentally friendly planetary science meetings.

\vspace{-.1cm}
\subsection{In-Person Meetings}
In-person meetings, compared to fully virtual meetings, are widely considered to have professional benefits, such as new collaborations forming from random encounters at the conference venue, early-career scientist finding job opportunities from personal interaction or notice boards, or eliminating time-difference issues inherent to virtual meetings. One, and evidently the most preferred option for the future American Astronomical Society's Division for Planetary Sciences (DPS) meeting would be to hold them the same way they were held in the past, with both the presenters and the audience present in a single meeting location. While it is the least environmentally friendly option, there are various ways to reduce  the negative environmental impact of the in-person meetings. Some options are:

\

{\bf 1. Strategic planning of meeting locations to minimize travel.} Using current data on  DPS membership, and assuming travel by air beyond 250 km, we find that DPS meetings in major cities within the contiguous US (and Canada) result in significantly lower flight miles, and consequently GHG emissions\footnote{Deriving the GHG emissions for a specific flight trip is a multi-parameter problem of which details can be read for example at www.icao.int/environmental-protection/Carbonoffset/Pages/default.aspx. See also \citet{ste20}.}, than those overseas, including Hawaii (Table \ref{tab:travel}).  A disproportionate fraction of GHG emissions associated with the meeting come from attendees taking intercontinental flights, which includes a large majority of DPS members if the meeting location is outside of North America. 

\begin{table}[htb]

\caption{Estimated distance of air travel assuming all 1184 DPS members attending a meeting in the given location. The calculated total air travel in km is obtained by adding great-circle distances between the venue and the city of record (or a nearby larger one) for all attendees travelling further than 250 km. Fourth and fifth column state the percentage of attendees traveling further than 5000 km, and the fraction of the value in the second column contributed by those attendees, respectively.}
\label{tab:travel}
\begin{center}
\begin{tabular}{ |c|c|c|c|c| } 
 \hline
 Location & Total air travel & People $<$250~km & People $>$5000~km & Travel $>$5000~km \\ 
 \hline
 Chicago & $3.1 \times 10^6$~km & 3\% &  14\% & 45\%\\ 
 Denver & $3.1 \times 10^6$~km & 5\% & 14\% & 46\% \\
 Honolulu & $7.7 \times10^6$~km & 1.5\% & 66\% & 79\% \\
London, ON & $3.2 \times 10^6$~km  & 1\% &  14\% & 42\% \\
Los Angeles & $3.6 \times 10^6$~km  & 13\% & 12\% &  39\% \\  
Paris & $8.5 \times 10^6$~km & 1\% & 93\% & 99\% \\
Washington & $3.3 \times 10^6$~km & 17\% & 14\% & 42\% \\
 \hline
\end{tabular}
\end{center}
\end{table}

\vspace{-.75cm}
Given the climate impact of intercontinental travel, future meetings outside of contiguous US (and Canada) would inevitably lead to more travel-related GHG emissions. Having meetings in the Baltimore/Washington DC area or in Pasadena would reduce the number of attendees who need to travel by air; meetings in Denver or Chicago would have fewer local attendees but require less travel on average for those coming from elsewhere in the US. Avoiding meetings in overseas locations such as Hawaii, or in smaller cities with fewer direct flight options may offer clearest benefits and should be strongly considered. A potential drawback may be financial, as some of these locations may be more expensive than the average for a US location. Also, completely abandoning international meetings would curtail international cooperation, which is a core goal of the DPS, AAS, and other scientific societies. As there is a some overlap between the communities attending the DPS and European Planetary Science Conference (EPSC) meetings with a number of scientists attending both in a row, we recommend that a similar study, based on actual attendance as in the work of Klöver (2020), be conducted to evaluate and compare the overall travel carbon footprint of separate (e.g. 2018, Knoxville and Berlin) vs joint (e.g. 2019 Geneva) DPS-EPSC meetings. More critical consideration of meeting locations has already strong support among planetary scientists, as shown by the recent community survey conducted by the DPS committee (https://dps.aas.org/reports). The same survey reveals that simply having fewer meetings has the least community support of all options for GHG emission reductions.
\

{\bf 2. Carbon offsets} Currently, carbon offsets for AAS meetings are voluntary and relatively few attendees purchase them, as the funding agencies do not allow them as a reimbursable cost. Including carbon offsets in the registration fee would provide an effective way to accumulate enough funds, but not necessarily a fair one nor legal as long as the cost is not federally allowable. Therefore, NASA and the planetary science community should advocate the carbon offsets to become an allowable cost. Additionally, the communication of the effect of the carbon offsets and their benefits to the meeting attendants is often unclear and could be improved. In cases where conferences choose to include carbon offsets into the registration fee, we recommend that exceptions are made for attendees travelling by means other than air or for attendees whose home institutions or nations have their own carbon offset plans in force. 

\

{\bf 3. Changing other, non-travel-related aspects of the meeting.} Plastic and food waste also play a major role in the anthropogenic GHG emissions. Therefore, reducing excessive and unnecessary printed materials that will soon be disposed of, as well as food and disposable container waste could offer other ways of decreasing the carbon footprint of the meetings. In addition to already implemented improvements such as using a smartphone application instead of a paper-printed abstract book, using reusable cups, plates, and cutlery (or alternatively recyclable or compostable single-use containers) together with requesting waste-recycling possibility throughout the venue would be desirable at future meetings. While personal food choices are outside of the purview of the DPS or DPSEAS, providing vegan or vegetarian food as a ``default'' option for conference food/snacks would be more environmentally friendly (\citet{ucs20} and references therein). 
\vspace{-.1cm}
\subsection{Virtual-Only Meetings}

The COVID-19 pandemic in 2020 has forced scientists to move as much of their in-person activities as possible online. This ongoing (at the time of writing) unplanned experiment has so far had varied results, but has shown that many aspects of lectures and meetings can be done remotely, if necessary. The AAS 2020 summer meeting had a virtual-only format, and as such, acted as a ``dress-rehearsal" for other virtual meetings. Additionally, the AAS's Division on Dynamical Astronomy virtual-only meeting in August is using a different format of asynchronous talks followed with real-time questioning of speakers, while the EPSC 2020 in September will be using a combination of synchronous and asynchronous presentations. These events will test concrete solutions to both the technological and intrinsic challenges of the virtual-only meetings.

\

While there would be some GHG emissions associated with computer hardware and network usage by both by the organizers and the attendees, a fully virtual meeting would generate zero travel-related GHG emissions. The shortcomings include technical limitations on participation (although these may not be intrinsic to the virtual format) or scheduling problems related to the spread of attendees over many time zones. We will not discuss the practical aspects of giving and following virtual presentations, as the upcoming virtual meetings may make that question moot, or uncover new, unforeseen issues. For an overview of implementing a virtual conference, see the report of the ACM Presidential Task Force on What Conferences Can Do to Replace Face to Face Meetings \citep{lop20} or the practical guide/white paper "A Nearly Carbon-Neutral Conference Model" \citep{hil20} originally drafted by Prof. Ken Hiltner (UCSB). 


\

Spontaneous interaction that is possible at in-person meetings but not at virtual ones is often anecdotally cited as a reason for keeping the traditional format. Attending meetings early in their career has certainly been beneficial to those among the authors who are now mid-career or senior researchers. Established scientists already have networks in place while early-career researchers new to the field have yet to meet many of their colleagues. Future changes in meeting format may bring changes to social dynamics, but that is outside of the scope of this White Paper.

\vspace{-.2cm}
\subsection{Hybrid Meetings}

Hybrid meetings would have some attendees present in person at a conference site (or multiple sites) while other attendees would be presenting and following presentations remotely. This approach has some benefits of both in-person and virtual meetings, but also presents some new challenges. The technical set-up must be such that remote participants can have a good view of in-person presentations, and also must be able to ask questions. The practical implementation requires further work but would likely overlap with solutions presented above in Section 2.2.

\

The exact financial structure of hybrid meetings may require some experimentation before the costs are well understood and planning is routine. Remote participants should pay a lower registration fee, as they are not making use of coffee, snacks or the venue overall. Fees would still need to cover most of audio-visual costs associated with remote presentations and viewing. Transitioning from in-person to hybrid meetings may require some flexibility with meeting budgeting on behalf of the DPS and the parent society.

\

{\bf A clear benefit of hybrid meetings is that they can greatly reduce travel time and cost, while still allowing some in-person interaction.} Hybrid meetings would offer new opportunities to members who cannot attend in person because of other work or family commitments, but would still like to take part in science sessions. Lower costs of online attendance would also open up the meetings to  researchers with less robust travel budgets. This could bring a significant increase in virtual attendance by students and also scientists from lower-income countries.

\

{\bf A special case for a two-location hybrid meeting can be made for joint DPS-EPSC meetings.} While there would be some issues related to the timezone (preferring a East Coast location for the DPS), there could be at least some sessions held in two rooms at two continents, with the presenters from the other location shown on the big screen. While this solution (DPS and EPSC in the same week, with some shared sessions) would not preserve all of the benefits of the joint meetings, it would greatly reduce GHG emissions due to travel of two sets of regular attendees.   

\vspace{-.2cm}
\section{Other Professional Travel}

Some forms of work-related travel are essential, including on-site telescope observations (at instruments which don't have a remote-observing capability) and fieldwork. Other forms of non-conference work-related travel may be easier to avoid and replace by remote work, such as those due to review panels and colloquia. NASA and the NSF have already organized remote and hybrid review panels, and remote is the standard form for review panels during the 2020 COVID-19 pandemic. While review panels are held by funding agencies and their contractors, the community still has an influence on their operation by volunteering or agreeing to reviewer duties. Academic department colloquia usually involve a visit by a scientist during which they would give one (or sometimes two) lectures at the host institution, and also meet with members of the host institution, typically over  1 or 2 days. If dozens of departments are to invite a speaker every week of the academic year, the total amount of travel is comparable to a large annual conference the size of DPS. However, unlike at a conference, where many people travel once, senior scientist may give many colloquia in one year, while most junior colleagues are invited much less often. 

\


{\bf A large reduction in the amount of long-distance travel for invited speakers to present in-person colloquia may be the ``low-hanging fruit'' of making planetary science more sustainable.} This can be accomplished by two methods: remote colloquia and coordinated visits. Remote colloquia are organized by many institutions and also offer an opportunity to reach audiences beyond the host institution through colloquium broadcasting or recording. Alternatively, colloquium committees of nearby institutions could coordinate speakers, providing 2 -- 4 colloquia for little more GHG costs than one colloquium under the model currently practiced by many financially-strong departments, and significant reduction in expenses per talk. While coordinated visits have a nontrivial GHG footprint, they allow the types of in-person interactions that cannot be well produced remotely. Both remote talks and coordinated travel lower the costs of colloquia, which may also result in more junior researchers invited to give talks. 


\

{\bf It is of paramount importance that the measures taken to reduce the planetary science community's carbon footprint should not be having unreasonable negative effects on the professional climate.} The scientific community should strive to ensure that the implementation of reductions in travel and other actions recommended in this white paper should not cause disproportionately negative impact on those in the profession who have less authority or resources, such as early-career scientists, and those who are already under-represented in the field (see Rivera-Valentín et al.). Equity, diversity, inclusion, and accessibility are critical steps to sustainable planetary science.


\vspace{-0.2cm}

\section{Other Activities}

Planetary science unavoidably requires offices, labs, computational resources, and large infrastructures. The infrastructure at the host institution is usually not under control of the individual scientists, but the community as whole can engage in efforts to influence host institutions and infrastructure operators to decrease the environmental footprint of their operations. Institutions and agencies could require more strongly environmental responsibility similar to requiring professional behavior from their staff or grant recipients. This requires great social reconstruction, but is crucial for the change of culture on the grass-root level. The following actions can be proposed:

\begin{itemize}
\item \textbf{Energy use optimization}: Investing into energy-efficient buildings (e.g., thermal isolation, heat recovery from ventilation air or cooling circuits, motion sensor-controlled lighting, LED lighting), energy-efficient equipment (e.g., low-power electric equipment, computers), and opting for renewable energy sources (e.g., renewable energy-based electricity contracts, on-site solar power generation for electricity and hot water production, heat pumps, heat recycling withing buildings) when possible. This is particularly crucial for facilities hosting supercomputers; according to \citet{ste20}, the majority of GHG emissions in astronomy are due to supercomputer usage.
\item \textbf{Consumption moderation}: Examples include but are not limited to reduction in use of paper and other resources or raw materials, moderating use of office space air conditioning, and rain water recovery and reuse (e.g., for irrigation).
\item \textbf{Waste management}: Environmentally friendly waste management maximizing recycling (including electronics, batteries, biowaste), reuse (e.g., donation of excess equipment), and adequate problematic waste handling. Provide easy access to appropriate receptacles.
\item \textbf{Transportation}: Use of alternate fuels in institution/campus vehicle fleets (waste-based biogas/biodiesel/ethanol, hybrid/electric vehicles), promoting carpooling or use of bicycles, and allowing remote work as often as possible.
\end{itemize}
A possible way forward in implementing these steps can be by formulating a community environmental policy. The universities, employers, and infrastructure operators can then be asked on behalf of the whole community to implement changes in line with the policy.



\

Institutions could also take a more active stance on positive change through education. The white Paper by \citet{wil19} gives an overview on how astronomers can embed climate change engagement in teaching and research. For planetary scientists, Earth and environmental sciences are often relatively close to their field of expertise, but few receive funding to study climate change, and thus they are not conflicted on the topic. Climate change could be embedded in the curricula of fundamental planetary science courses where applicable.

\vspace{-0.2cm}

\section{Recommendations}
Climate change will be one of the defining aspects of the next decade, which will touch all human lives, including the scientific communities. Leadership and cultural change are crucial to mitigate the damage that the AGW will cause not only to the scientific community but all of human kind, and the actions must be taken now. While the following recommendations could cause some inconvenience, readiness to face inconvenience must be accepted by the community in order to reach the goal of decreasing GHG emissions. 
We recommend the funding agencies as well as scientific institutions to:

\ 

{\bf Assess the carbon footprint of the planetary science community in the US, including emissions from travel, supercomputing and other facilities, as well as space programs.} This step is crucial to establish a realistic picture of the current situation and a reference for the effects of the mitigation measures.

\

{\bf Encourage organizing of virtual or hybrid conferences and workshops and virtual review panels.} We recommend that every effort should be made to make all planetary science meetings fully virtual or hybrid meetings with a virtual component at least on alternating years. Now that the technology for virtual presentation and participation is widely available, and with full understanding of the climate impact of long-range air travel, it is irresponsible to stay restricted to the old paradigms of scientific meetings. Hybrid meetings would ensure that there is a continuity with past practices and that there are opportunities for in-person interaction but would also make it possible for people who would otherwise be unable to attend in person due to time or budget constraints to take part. Invitees for review panels should agree to participate remotely especially when the ratio of overall travel distance to meeting duration is large. \\

%

\

{\bf Make the carbon offsets an allowable and encouraged cost, unless/until the full global costs of GHGs is priced into travel via carbon taxes, cap-and-trade or other mechanisms.} This will encourage a greater numbers of community members to take on this responsibility. We recommend that carbon offsets are made an option and strongly encouraged by agencies and institutions to be paid by anyone traveling to a non-local scientific meeting. Appropriate companies providing the implementation of carbon-offsetting services must be assessed.

\

{\bf Require environmental responsibility from the scientific community.} Similar to funding agencies and institutions requiring their staff, meeting attendants, and grant recipients to take responsibility in professional climate, diversity, and zero-tolerance for harassment, being responsible in environmental issues is equally important for the continuation of the well-being of the members of the community. A cultural change in  attitude is crucial for grass-root level change.

\vspace{-.1cm}

\bibliographystyle{apalike}

\end{document}